\documentclass[prl,twocolumn,showpacs,preprintnumbers,superscriptaddress,floatfix,nofootinbib,reprint]{revtex4-1}

\usepackage[left]{lineno}

\usepackage{amsfonts}
\usepackage{multirow}
\usepackage{subfigure}

\usepackage{graphicx,,booktabs}
\usepackage{amssymb}
\usepackage{amsmath,txfonts,ulem}
\usepackage{graphicx}
\usepackage{dcolumn}
\usepackage{color}
\usepackage{bm}
\usepackage{ulem}

\usepackage{enumerate}

\usepackage{appendix}
\usepackage[colorlinks,
 citecolor=blue,
 anchorcolor=green,
 menucolor=orange,
 linkcolor=blue,
 filecolor=red,
 runcolor=pink,
 urlcolor=blue,
 frenchlinks=red]{hyperref} 

\allowdisplaybreaks[4]

\begin{document}

\title{\boldmath Self-Cancelation of Coherent Synchrotron Radiation Kicks Using\\ a Non-Symmetric  S-shape Four-Bend  Chicane  
    } 

\author{ Fancong Zeng}
\affiliation{ Key Laboratory of Particle Acceleration Physics and Technology, Institute of High Energy Physics, Chinese Academy of Sciences, and University of Chinese Academy of Sciences, Beijing 100049, China } 
  
\author{Yi Jiao}
\email{jiaoyi@ihep.ac.cn}
\affiliation{ Key Laboratory of Particle Acceleration Physics and Technology, Institute of High Energy Physics, Chinese Academy of Sciences, and University of Chinese Academy of Sciences, Beijing 100049, China }

\author{Weihang Liu}
\affiliation{China Spallation Neutron Source, Institute of High Energy Physics,
Chinese Academy of Sciences, Dongguan 523803, China }

\author{Cheng-Ying Tsai}
\affiliation{ School of Electrical and Electronic Engineering,
Huazhong University of Science and Technology, Wuhan 430074, China }

\date{\today}
\begin{abstract} 
High peak current electron beams are essential for x-ray free-electron lasers (FELs), and generally realized through multi-stage compression with symmetric C-shape four-bend chicanes. However, the coherent synchrotron radiations (CSR), emitted for wavelengths longer than or comparable to the length of the electron bunch during the compression, may degrade the beam quality and finally affect the FEL performance. In this Letter, we show that zero net CSR kick cannot be achieved in a symmetric C-chicane by using an explicit point-kick analysis of the CSR effects, which is responsible for significant emittance growth when pursuing a peak current of $\gtrsim$ 10 kiloamperes. A four-bend chicane with non-symmetric S-shape geometry that can self-cancel the CSR kicks is proposed to effectively suppress the emittance growth. Compared to the symmetric C-chicane, beams with three times higher peak current and similar emittance growth can be achieved with the S-chicane for typical FEL operation parameters. We believe that this study provides a viable way of producing high quality and short intense electron beams, benefiting future development of FELs and other types of accelerator-based scientific facilities.

\end{abstract}

\maketitle 

The remarkable capabilities of x-ray free electron lasers (FELs) are of paramount importance to scientific community \cite{cai-1, cai-2, cai-3}, as FEL can generate intense, subterawatt-femtosecond (fs) x-ray pulses, enabling groundbreaking experiments and discoveries across a wide range of research fields. 
 Studies showed that users shall greatly benefit from having at least 10 to 100 times more coherent photons within a comparable or shorter pulse duration than typical existing x-ray FELs \cite{cai-4,cai-5,cai-6,cai-7}. 
For the electron beam,  a final peak current of up to 10 kiloamperes (kA) at the \text{10 GeV} energy level is required, characterized by a sufficiently small energy spread (on the order of 0.01\%) and low normalized transverse emittance (e.g., $1 ~ \mu \text{m-rad}$ or less) \cite{Tsai:2018wjs}.

\begin{figure}[h]
 	\centering
 	\includegraphics[scale=0.14]{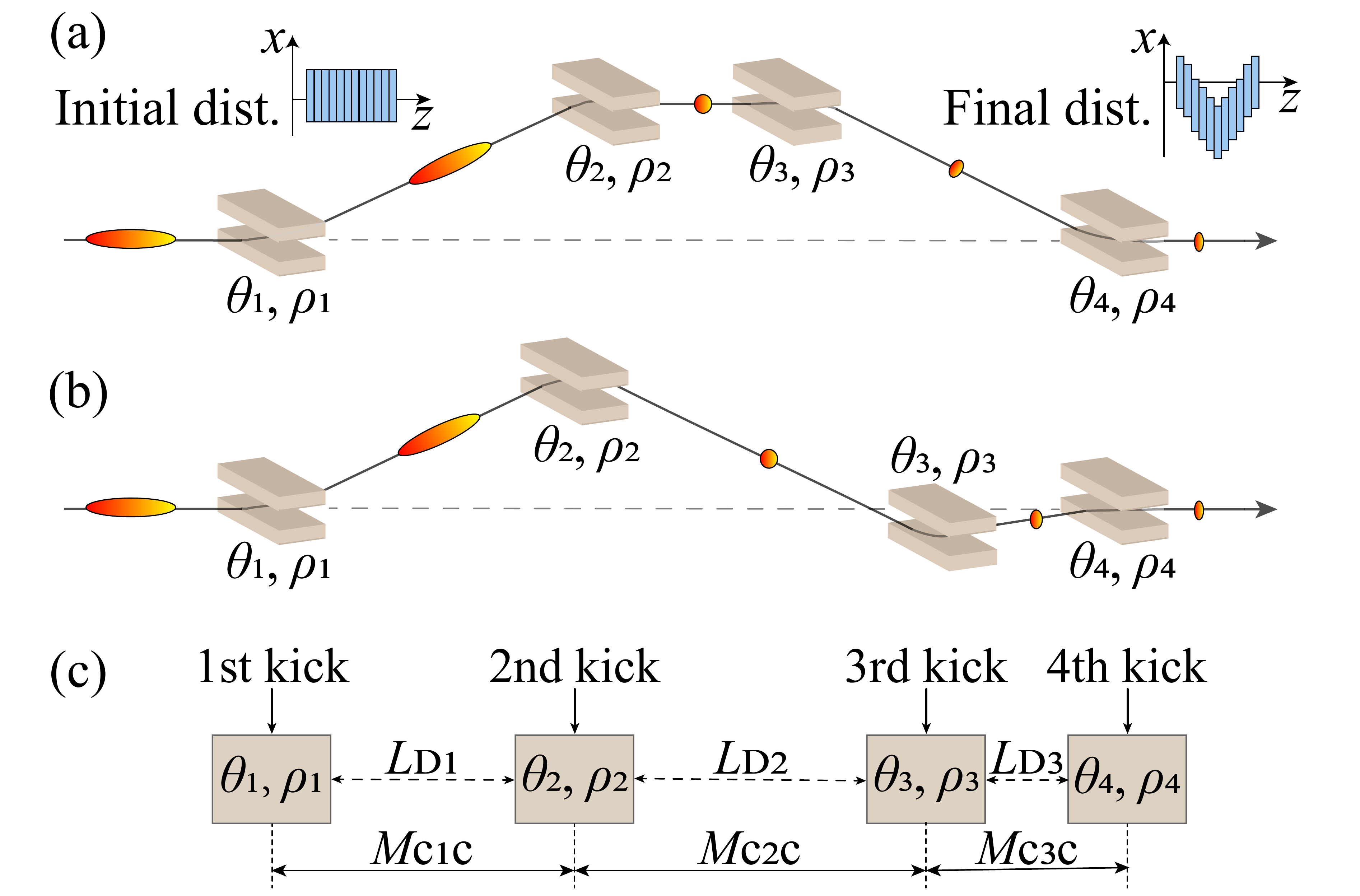}
 	\caption{ (a) Schematic of a symmetric C-chicane (not to scale).  (b) Schematic of a general four-bend chicane (not to scale). (c) Schematic of the point-kick analysis of the CSR effects. The arrows point to the centers of the bends.  $M_{\text{c}i\text{c}}~(i=1,2,3)$ represent the transfer matrices from the center of $i$th bend to the center of $(i+1)$th.  }
 	\label{fig:lattice2} 
 \end{figure}

Most of the existing FEL facilities operate with a peak current of typically 3-5 kA via multi-stage bunch compression \cite{FLASH,Allaria2012,Emma:2010eov,Ishikawa,Kang,weise,Milne}.
To a certain extent the principle of bunch compression is similar to that of the chirped pulse amplification technique \cite{Strickland:1985gxr}: 
an electron beam is imprinted with an energy chirp by an RF cavity, and then passes through a longitudinal dispersive section where particles with higher (lower) energy at the tail (head) of the bunch travel a shorter (longer) path \cite{Bartosik:2022ebg}. 
The most commonly employed dispersive section is symmetric C-chicane, composed of only four identical bending magnets (see Fig.  \ref{fig:lattice2}{\color{blue}(a)}), which is straightforward to implement and highly effective.
  
However, when bent, electrons produce broadband synchrotron radiations. These components, with wavelengths longer than or comparable to the bunch length, are called coherent synchrotron radiation (CSR).  
CSR establishes correlations between the tail and head particles of the bunch, potentially leading to collective effects. The shorter and more intense the bunch is, the stronger the CSR fields produced, and the more severe the CSR effects become. Especially when pursuing a 10 kA or even higher peak current in the final stage C-chicane compressor, the CSR effects may lead to significant longitudinal microbunching instability \cite{Huang:2002kp,Huang:2004ve,Tsai:2020ddc,Tsai:2017wyq,Roussel:2015vix,Tsai:2017hef,Venturini:2007zzb,Venturini:2007zzd,caity,Heifets:2002qt} and transverse emittance growth  \cite{Saldin:1996gs,Braun:2000dv,Braun:1999eq, Bane:2008zz}, and eventually degrade the FEL performance \cite{Hara:2016arh,Hara:2018qst}.

Regarding the undesired CSR-induced emittance growth in a chicane compressor, a number of suppression methods have been proposed in the past decades, including longitudinal pulse shaping \cite{Mitchell:2013tla, Penco:2014sza}, dispersion-based beam-tilt correction \cite{Guetg:2015gra, Guetg:2016sdy},  $\pi$ phase advance matching between adjacent achromats \cite{Dohlus:1998,DiMitri:2013qj,BC2design, Douglas:2014cia}, and initial Courant-Snyder (C-S) parameter scanning \cite{Hajima,Jing:2013cma}. 
However, the first three approaches are effective but require additional techniques or components, and the last one is itself of limited effect.  Another desirable approach is to explore different chicane geometries with an emphasis on CSR compensation for high-peak current compression, for instances non-symmetric C-shape and S-shape chicanes with four or more bends \cite{Loulergue:2001kh,Stulle:2007se,Stulle:2004tt,NLS,Williams:2010zzf,Khan:2022tkg,Bartolini:2009,Borland:2000ah,Bartolini:2012zza}. However,  this approach usually relies heavily on numerical simulations and parameter scans, and works on a case-by-case basis. It remains unclear whether there exist general guidelines for the design of a CSR-immune chicane compressor.
 
In this Letter we report a novel non-symmetric, S-shape, four-bend chicane that self-cancels the CSR kicks.
With a point-kick analysis and particle tracking simulations, we present the design criterion of the proposed S-chicane and show its excellent emittance preservation capability. By using such an S-chicane instead of a C-chicane, the CSR-induced emittance growth can be reduced to about one order of magnitude lower level in an almost the same bunch compression scenario, and moreover, about three times higher peak current can be produced for the same amount of emittance growth.  
 
For definiteness, we consider a general model of a four-bend chicane, sketched in Fig. \ref{fig:lattice2}{\color{blue}(b)}. The length of each bend is $L_\text{B}$, while the bending angles of the four bends $\theta_i~(i=1,2,3,4)$  are not required to be equal and the three drifts  $ L_{\text{D}j}~(j=1,2,3)$ with $ L_{\text{D}j} \gg L_\text{B}   $ are not pre-constrained to any specific relation.

For such a chicane, the achromatic condition and beam coaxial condition are not naturally satisfied. Nevertheless, these two conditions share the same requirements \cite{Material}, 
\begin{equation}\label{eq:achromatic-condition} 
\begin{aligned}
&1+q_2+q_3+q_4=0,\\
&1+\ell_2 (1+q_2 )+\ell_3 (1+q_2+q_3 )=0,	
\end{aligned}
\end{equation}	
where normalized bending angles, $q_i=\theta_i/\theta_1 $  $(i=1,2,3,4)$, and normalized drift lengths,  $ \ell_j=L_{\text{D}j}/L_{\text{D}1}~(j=1,2,3)$, are introduced. 

With the aid of Eq.  \eqref{eq:achromatic-condition}, the total length and longitudinal dispersion of the chicane,  $L_{\text{tot}}$ and $R_{56}^{s_0\to s_f }$,   can be expressed as
 \begin{equation} \label{eq:tot+R56}
 L_{\text{tot}} =  (1+\ell_2+\ell_3 ) L_{\text{D1}},~ R_{56}^{s_0\to s_f} = L_{\text{D1}} \theta_1^2  (q_2+\ell_3 q_3 q_4  ) . 
\end{equation}
In the following analysis, linear compression is considered, with the compression factor $C\equiv \sigma_{z0}/ \sigma_{zf}  =1/(1+hR_{56}^{s_0\to s_f } ) $, where  $ \sigma_{z0}, \sigma_{zf}$  are the initial and final bunch lengths respectively, and $h$  is the initial energy chirp.

This model can easily degenerate to that of a symmetric C-chicane with $q_2=q_3=-1,q_4=1$ and $\ell_3=1$. 
On the other hand, this model has a larger number of adjustable parameters compared to a symmetric C-chicane which, as shown below, provides sufficient knobs for self-cancelation of the CSR kicks. Besides, it keeps a simple layout with only four bends, and is almost transverse dispersion free at all orders \cite{Loulergue:2001kh, Stulle:2004tt,Stulle:2007se}, which benefits beam quality preservation as in a symmetric C-chicane.

  We then derive the analytical expressions of net CSR kick at the chicane exit and  the CSR cancelation conditions using the point-kick analysis \cite{Jiao:2014gja}, where the cumulative steady-state CSR (denoted as ss-CSR) effects along the particle trajectory in a bend can be simplified to a point kick at the center of the bend. After experiencing a CSR kick, the particle will have a horizontal coordinate deviation $\Delta X_{\text{CSR}}$  and an additional energy deviation $\Delta \delta_{\text{CSR}}$ relative to synchronous particle \cite{Jiao:2014gja},
\begin{equation} \label{eq:Xk}
\begin{aligned} 
\Delta X_{\text{CSR}}&= 	\left[   \begin{array}{c  }
\Delta 	x_{\text{CSR}}  \\
\Delta 	x_{\text{CSR}}^\prime
	\end{array}   \right]=
	\left[   \begin{array}{c  }
	\kappa	\rho^{4/3}  (\theta \cos(\theta /2)-2 \sin(\theta /2) )  \\
		\sin(\theta /2) (2 \delta +\kappa \rho^{1/3} \theta   )
	\end{array}   \right] ,\\
 \Delta  \delta_{\text{CSR}} & =  \kappa\rho^{1/3} \theta, 
\end{aligned} 
\end{equation} 
where $\delta$ is the energy deviation before experiencing this CSR kick, $\rho$ is the beam bending radius, and the coefficient $\kappa$ is of the form   
 \begin{equation}\label{eq:int2}
\kappa=0.2459\frac{N_b r_e}{\gamma \sigma_z^{4/3}}, 
\end{equation}
with $N_b$ is the electron population, $r_e$ is the classical electron radius, $\gamma$ is the relativistic Lorentz factor, and $\sigma_z$ is the rms electron bunch length.

It should be noted that the bunch length is varying along the path length, and $\kappa$ is actually not constant in a bend. For simplicity, we assume a constant, lumped $\kappa$ for each CSR kick, corresponding to a constant bunch length within each bend, namely a constant $\sigma_{zi}~(i=1,2,3,4)$ in the $i$th bend  \cite{Mitri:2015,DiMitri:2016gia,Zhang:2023cgl}.
This assumption, despite sacrificing precision to some degree, preserves the main physics of compression, and especially allows an explicit analysis of CSR effects. Specifically, for the second (third) CSR kick we use the bunch length at the center of the second (third) bend; and for the first and the fourth kicks we use the bunch lengths at the entrance and exit of the bends respectively,
 i.e., $\sigma_{z1}=\sigma_{z0}$  and $    \sigma_{z4}=\sigma_{zf}$,  which allows for a simple relation $\kappa_4= \kappa_1 C^{4/3}$.

Through a straightforward matrix multiplication, the net CSR kick,  $(\Delta x_{\text{CSR,f }},\Delta x_{\text{CSR,f}}^{\prime} )^T$, can be obtained after successively counting the four CSR kicks and the three sections between the centers of adjacent bends (see Fig. \ref{fig:lattice2}{\color{blue}(c)}) \cite{Material}. This net kick is related to the final emittance by 
 \begin{equation}\begin{aligned}
&\varepsilon_{nf}^2=\varepsilon_{n0}^2+\\
&\varepsilon_{n0}\left(\beta_x\left\langle\Delta x_{\text{CSR,f}}^{\prime2}\right\rangle+2\alpha_x\left\langle\Delta x_{\text{CSR,f}}\Delta x_{\text{CSR,f}}^\prime\right\rangle+\gamma_x\left\langle\Delta x_{\text{CSR,f}}^2\right\rangle\right) ,
\end{aligned}
\end{equation}
where $ \varepsilon_{n0}$ is the unperturbed geometric emittance, and $\alpha_x,\beta_x,\gamma_x$ are the C-S functions at the chicane exit. 
If the net CSR kick is equal to \textit{zero}, in principle, there will be no CSR-induced emittance growth. Combining Eq. \eqref{eq:achromatic-condition}, the CSR cancelation conditions can be obtained \cite{Material},  
 \begin{equation}  \label{eq:condition3} 
 	\frac{ q_2  }{q_3 q_4 \ell_3 }  =\frac{\Lambda_3 }{\Lambda_1 },~~ 	\ell_2=  -\frac{1}{q_3} \frac{   (q_2+q_3)   \Lambda_1    + q_3 \Lambda_2  }{    \Lambda_1  +( 1+q_2)  \Lambda_2 } ,
 \end{equation} 
in which  $\Lambda_1=\sigma_{z1}^{-4/3}  +\sigma_{z2}^{-4/3} q_2^{2/3}$,  $\Lambda_2=\sigma_{z2}^{-4/3}   q_2^{2/3}  +\sigma_{z3}^{-4/3} q_3^{2/3}$ and $\Lambda_3=\sigma_{z3}^{-4/3} q_3^{2/3} +\sigma_{z4}^{-4/3} q_4^{2/3}$ 
are all positive. The CSR cancelation conditions appear complicated, with the requirements on bending angles and drift lengths coupled together. Nevertheless, with Eq.
 \eqref{eq:condition3} it is enough to investigate the feasibility of canceling the CSR kicks in a four-bend chicane.

First, for the widely used symmetric C-chicane, by substituting the relations $q_2=q_3=-1,q_4=1$ and  $\ell_3=1$  into Eq. \eqref{eq:condition3}, one can find that $ \ell_2$ is an impractical negative value (it can be realized with a focusing section \cite{fancong:2023} but beyond the scope of this Letter). Second, for the case of rotationally symmetric (S-shape) layout with $q_2=-q_3,q_4=-1$  and $\ell_3=1$, it is also impossible to realize  \textit{zero} net CSR kick.
At last, for a general case without any symmetric geometry constraints, an analysis for all possible scenarios turns out that to reach \textit{zero} net CSR kick, one needs to set the bending angles as (see Ref. \cite{Material} for details)
\begin{equation}\label{eq:qn}
q_1=1, ~ q_2<-1, ~ q_3>0  \text{ and }  q_4<0.
\end{equation}
This suggests that the unique way of canceling the CSR kicks in a four-bend chicane is to reshape the chicane to a non-symmetric S-shape layout, and simultaneously, make the chicane parameters satisfy the requirements in Eq. \eqref{eq:condition3}.
It is worth mentioning that the concept of non-symmetric S-chicane design has been previously proposed and explored  \cite{Loulergue:2001kh,Stulle:2004tt}. Nevertheless, to the best of our knowledge, our study for the first time clearly signifies that using four bends is enough to completely cancel the CSR kicks. Next we will present in the following a specific design criterion of such an S-chicane.

\begin{figure}
\centering
\includegraphics[scale=0.32 ]{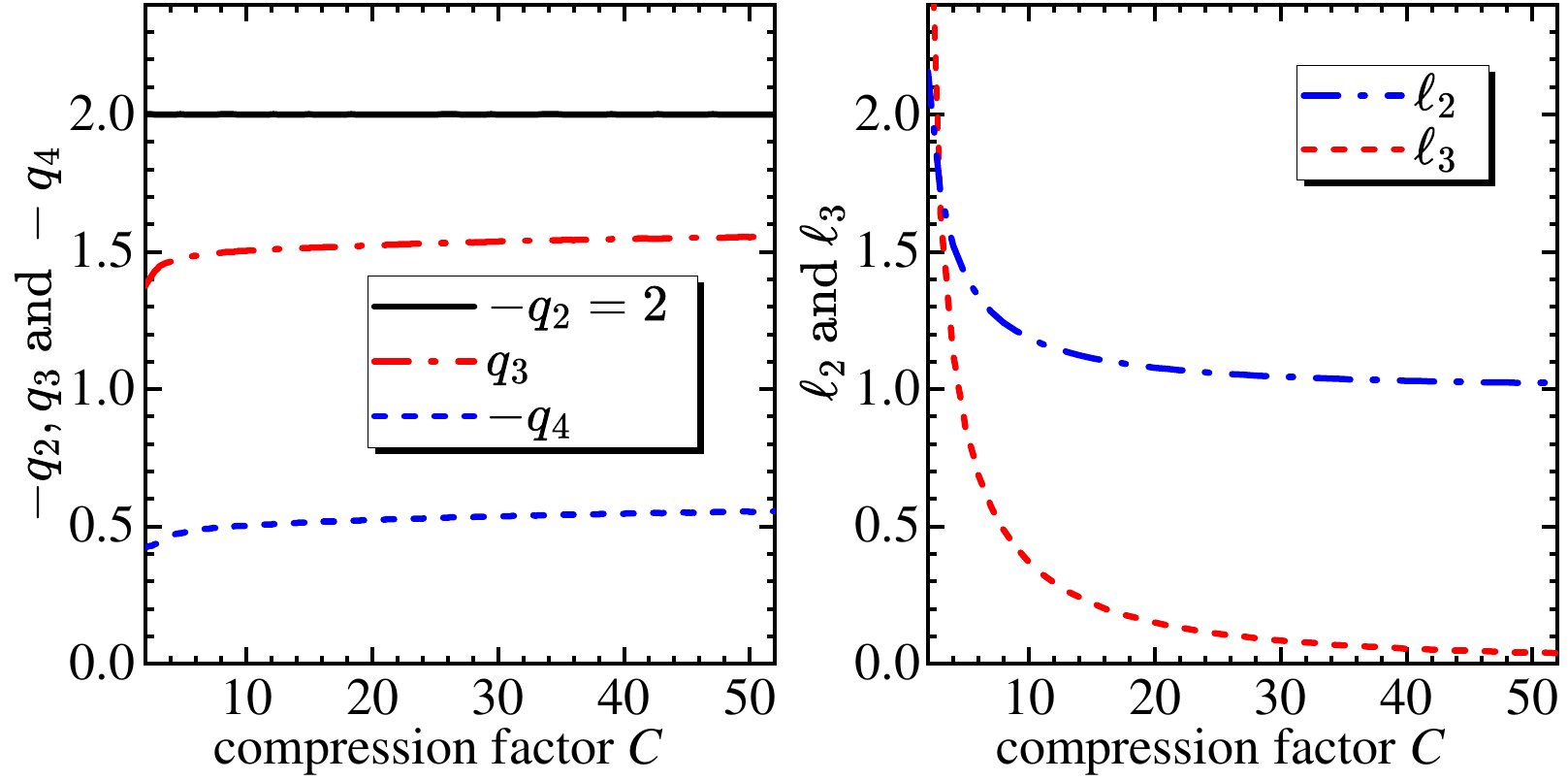}  
\caption{The values of $q_3, q_4, \ell_2,\ell_3$ versus compression factor  $C$   by solving Eqs. \eqref{eq:achromatic-condition} and  \eqref{eq:condition3} with   $q_2 \equiv -2$.  }
\label{fig:CSR} 
\end{figure}

From the above analysis, there are a total of  \textit{six} variables,  i.e., $q_2,q_3,q_4,\ell_2,\ell_3$ and $C$; whereas the requirements in Eqs.\eqref{eq:achromatic-condition} and  \eqref{eq:condition3}  impose  \textit{four} constraints. To obtain a dependency relationship between the chicane parameters and compression factor $C$, it is necessary to fix \textit{one} variable. There are many options for this. Here we somewhat arbitrarily choose a case with $q_2 \equiv -2$, which resembles the angle setting of a symmetric S-chicane \cite{Loulergue:2001kh,NLS,Williams:2010zzf,Bartolini:2009}. In this case, the required bending angles and drift lengths with different compression factors can be numerically solved from Eqs. \eqref{eq:achromatic-condition} and  \eqref{eq:condition3}, with the results presented in Fig. \ref{fig:CSR}. It can be seen that the desired $(q_3,q_4,\ell_2,\ell_3)$ vary with $C$. Especially, a larger compression factor $C$ leads to a smaller $\ell_3$, i.e., a shorter distance between the last two bends. From a practical point of view, to avoid the distance being excessively small, one can limit $\ell_3$ to not less than 0.2, corresponding to a compression factor of not larger than 16, which can meet the requirement on compression factor of a single chicane in most FELs \cite{FLASH,Allaria2012,Emma:2010eov,Ishikawa,Kang,weise,Milne}. 

To form such an S-chicane from a widely used symmetric C-chicane, it is necessary to power the four bends independently and install the two inner bends on remotely movable horizontal stages \cite{Vogt:2021cpo, Arthur:2002ap}. In practice, sometimes it may need to change the compression factor of a chicane, while keeping the total length unchanged. To achieve this in the proposed S-chicane, the parameters $(q_3,q_4,\ell_2,\ell_3)$ are needed to adjusted according to the results in Fig. \ref{fig:CSR} to ensure the CSR cancelation and achromatic conditions; and following Eq. \eqref{eq:tot+R56}, the longitudinal dispersion  can be adjusted or kept unchanged, by choosing appropriate values of $L_{\text{D1}}, \theta_1$  and $\theta_2\equiv -2\theta_1$. These can be achieved by programming the driving currents of the four bends and adjusting the positions of two inner bends.

Next the CSR suppression efficiency of the proposed S-chicane is tested with a numerical example, where a Gaussian electron bunch is launched. Typical initial beam parameters are used, with a normalized emittance of 0.9 $\mu$m-rad, a beam energy of 3 GeV, an uncorrelated energy spread of 0.01\%, a bunch charge of 300 pC, and a bunch length of  100 $\mu$m. A chicane with a total length of \text{20 m} and a longitudinal dispersion of  $-39$ mm is considered. According to the linear compression relation the initial energy chirp $h$ is set to $23.12\text{ m}^{-1}$, to achieve a compression factor of 10 and a peak current of \text{3.4 kA}  at the chicane exit. The other chicane parameters are listed in Table \ref{Table-S-chicane}.  
  
The particle tracking in the presence of the CSR effects is implemented with the ELEGANT program \cite{Borland:2000gvh,Borland:2001xua}. In the simulation, two cases, with the ss-CSR only and a full CSR model, are considered. The former is used to verify the results from the above point-kick analysis. Considering that the net CSR kick would not be exactly zero, in each case initial C-S parameter scanning \cite{Hajima} is employed to find the smallest possible emittance growth. The simulation results are also listed in Table \ref{Table-S-chicane}. In the case with the ss-CSR only, the relative emittance growth  is only about 1\%, indicating a successful cancelation of the CSR kicks. In the case with a full CSR model, the relative emittance growth rises to 5.5\%. Further analysis indicates that the additional emittance growth is mainly attributed to the CSR wake in the drift spaces following bends \cite{Saldin:1996gs,Stupakov:2002xs}.

For comparison, we simulate the CSR-induced emittance growth in a symmetric C-chicane, where the same initial beam parameters and compression target, namely the same $L_{\text{tot}},  R_{56}^{s_0\to s_f}$ and $C$, are used, and initial C-S parameter scanning is also performed. Detailed parameters and simulation results are also listed in Table \ref{Table-S-chicane}. The relative emittance growth is about 14.5\% (55.6\%) in the case with the ss-CSR only (a full CSR model). It can be seen from Table  \ref{Table-S-chicane} that the proposed S-chicane has obviously larger bending angles in the two inner bends as well as a larger total bending angle (absolute value), suggesting stronger CSR than in the C-chicane. However, the proposed S-chicane allows about one order of magnitude lower relative emittance growth, because of the self-cancelation of the CSR kicks. In addition, for both chicanes the higher-order longitudinal dispersions $T_{566}$ are the same, and the microbunching instability gains remain in the same order of magnitude \cite{Liu:2024mrv}.

 \begin{table}
\caption{Parameters of a symmetric C-chicane and non-symmetric S-chicane and simulation results of the relative emittance growth.  
    } \centering
\begin{tabular}{lccc  }
\hline	\hline\noalign{\smallskip}
\multirow{2}*{Parameter} 	&\multirow{2}*{ Symbol}   & ~ symmetric~  &   ~non-symmetric~      \\  
	&    &  C-chicane  &          S-chicane    \\ \hline	\noalign{\smallskip} 
	Length of    dipoles (m)  &$ L_{\text{B} } $   &	 0.81 &	 0.81   \\  
		\noalign{\smallskip} 	 	 
Angle of   1st   dipole ($ ^{\circ} $) &$ \theta_{1} $    & 2.68 &      2.80 \\ \noalign{\smallskip}
Angle of  2nd dipole ($ ^{\circ} $)  &$ \theta_{2} $    & $ -2.68$  &$-5.60$  \\ \noalign{\smallskip}
Angle of  3rd dipole ($ ^{\circ} $) &$ \theta_{3} $    &  $-2.68$  &     4.21 \\ \noalign{\smallskip}
Angle of   4th  dipole ($ ^{\circ} $)  &$ \theta_{4} $ &  2.68 &    $-1.41$ \\ \noalign{\smallskip}
Length of   1st drift (m) &$ L_{\text{D1}}$   & 3.99  & 6.87 \\  \noalign{\smallskip}
Length of    2nd drift  (m)  &$ L_{\text{D2}} $   & 0.01   &   8.15  \\  \noalign{\smallskip}
Length of   3rd drift  (m)  &$ L_{\text{D3}}$   &3.99   &  1.74 \\  
\hline	  \noalign{\smallskip} 
 &  \multicolumn{3}{c}{  steady-state CSR included } \\   
 \hline	  \noalign{\smallskip} 
Emittance growth ratio & $\Delta \varepsilon_{n}/\varepsilon_{n0}  $  &14.5\% &1.0\%\\ 
  \hline \noalign{\smallskip} 
 & \multicolumn{3}{c}{a full   CSR model  included } \\ 
     \hline	  \noalign{\smallskip} 
Emittance growth ratio& $\Delta \varepsilon_{n}/\varepsilon_{n0}  $ &55.6\% & 5.5\%   \\   
\hline	\hline

\end{tabular} \label{Table-S-chicane}
\end{table} 

To test robustness of the proposed S-chicane, we conduct a scan of the parameters $(q_3,\ell_2)$  within a relative deviation range of $[-10\%, +15\%]$ from the analytically predicted values denoted as  $(q_3^*,\ell_2^*)$, where $(q_4, \ell_3)$ are adjusted accordingly to ensure the achromatic and coaxial conditions. 
The relative emittance growth remains at the same level, with the smallest relative emittance growth of 4.7\% at $(q_3/q_3^*, \ell_2/\ell_2^*) = (1.08, 0.99)$ in the presence of a full CSR model. This indicates that the CSR suppression efficiency is insensitive to the errors of these parameters, and the proposed S-chicane is robust in suppressing the CSR-induced emittance growth.
 
\begin{figure} 
\centering
\includegraphics[scale=0.38]{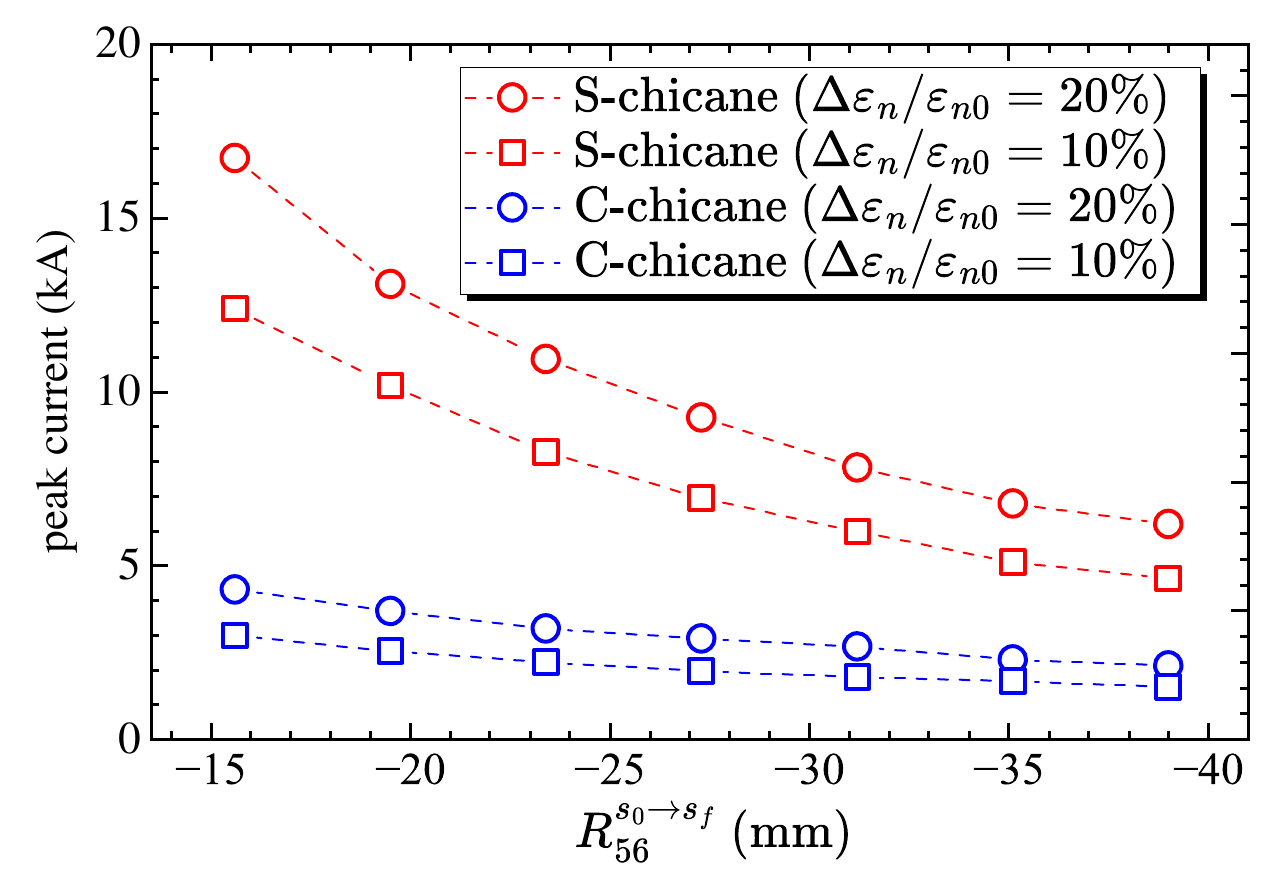}  
\caption{Peak current versus $R_{56}^{s_0\to s_f}$ for symmetric C-chicane (blue points) and non-symmetric S-chicane (red points) while  maintaining the CSR-induced  relative emittance growth to  10\% (square points) or 20\% (round points).
 }
\label{fig:3} 
\end{figure}

Because of the excellent emittance preservation capability, the proposed S-chicane could be an appealing option for the final stage compressor, where the CSR effects generally arise associated with the high-peak current compression. Here we make a preliminary investigation of the maximum available peak current of the proposed S-chicane. From current FEL design and operation experiences, e.g., \cite{3-BC,Loulergue:2001kh}, the final stage C-chicane compressor is usually designed with a smaller longitudinal dispersion, and simultaneously, with the peak current controlled to a moderate level, typically \text{3-5 kA}, so as to keep the CSR effects at an acceptable level. Similarly, in this test we fix the initial beam parameters and chicane settings as in    \text{Table \ref{Table-S-chicane}}, except for the bunch charge (and hence the target peak current) and bending angles (and hence the longitudinal dispersion). We then simulate the CSR-induced emittance growth in the presence of a full CSR model. The results are presented in Fig. \ref{fig:3}  (See Ref. \cite{Material} for further details). It appears that by using the proposed S-chicane rather than a symmetric C-chicane, about three times higher peak current can be allowed for the same amount of relative emittance growth. For instance, in the case with a longitudinal dispersion of $-15.6$ mm, it is feasible to produce a peak current of 3.0 kA (4.3 kA) by use of a C-chicane, while  a peak current of 12.4 kA (16.7 kA) with the proposed S-chicane, for a relative emittance growth of 10\% (20\%).

In summary, we have demonstrated the excellent emittance preservation capability of the proposed non-symmetric four-bend S-chicane when used to produce a kA level or even higher peak current. The underlying physics is to use a non-symmetric bend layout to cancel the non-symmetric CSR kicks in a chicane compressor. Such an S-chicane can be converted from a symmetric C-chicane with just a few modifications in the same total length. It would not be difficult for the design and optimization of experiments in existing FEL facilities. We believe that the study presented here and further investigations will make the non-symmetric four-bend S-chicane a viable way of producing high intense and high quality electron beams, benefiting future development of the FEL and other types of accelerator-based scientific facilities.

\begin{acknowledgements}
This work was supported by the National Natural Science Foundation of China (No. 12275284 and No. 12275094), the Fundamental Research Funds for the Central Universities (HUST) under Project No. 2021GCRC006, and National Key Research and Development Program of China (No. 2022YFA1603402).
\end{acknowledgements}

\end{document}